\documentclass[twoside]{dis08}
\usepackage[latin1]{inputenc}
\usepackage[dvips]{graphicx,epsfig,color}
\usepackage{wrapfig,rotating}
\usepackage{amssymb,amsmath,array}

\newcommand{\vl}{\pmb{l}}
\newcommand{\vk}{\pmb{k}}

\newcommand{\bea}{\begin{eqnarray}}
\newcommand{\eea}{\end{eqnarray}}

\newcommand{\beq}{\begin{equation}}
\newcommand{\eeq}{\end{equation}}
\newcommand{\beqar}[1]{\begin{eqnarray}\label{#1}}
\newcommand{\eeqar}{\end{eqnarray}}

\newcommand{\beeq}{\begin{eqnarray}}
\newcommand{\eeeq}{\end{eqnarray}}

\def\Ba{D}

\begin{document}
\title{Proton elastic impact factors for two, three, and four~gluons}

%***********************************************************************
% AUTHORS INFORMATION AREA
%***********************************************************************
\author{Leszek Motyka$^{1,2}$
%
% Optional short acknowledgment: remove next line if non-needed
\thanks{The support of the DFG grant no.\ SFB676 is gratefully acknowledged.}
%
% DO NOT MODIFY THE FOLLOWING '\vspace' ARGUMENT
\vspace{.3cm}\\
%
% Addresses and institutions (remove "1- " in case of a single institution)
1- II Institute for  Theoretical Physics, University of Hamburg,\\ 
Luruper Chaussee 149, D-22761, Germany \\[1mm]
2- Institute of Physics, Jagellonian University \\
Reymonta 4, 30-059 Krak\'{o}w, Poland
%
% Remove the next three lines in case of a single institution
%\vspace{.1cm}\\
%2- School of Second Author - Dept of Second Author \\
%Address of Second Author's school - Country of Second Author's school\\
}
%***********************************************************************
% END OF AUTHORS INFORMATION AREA

\maketitle

\begin{abstract}
In this talk \cite{url} we report on recent calculation of high energy baryon scattering amplitudes in QCD. Elastic baryon impact factors for two, three and four gluons are presented and their energy evolution is described that incorporates unitarity corrections. We find that the baryon couples directly to the BFKL Pomeron, to the BKP odderon and to a new state, a three-gluon BKP Pomeron. The new state may decay into four gluons with a new $3\to 4$ transition vertex. This vertex defines the transition amplitude of the three-gluon BKP Pomeron state into two BFKL Pomerons.
\end{abstract}

\section{Introduction}

The complete picture of high energy scattering of hadrons in QCD has to
incorporate effects of multiple scattering. In particular, when those 
effects are properly taken into account, they solve the problem of  
the rapid increase of Balitsky-Fadin-Kuraev-Lipatov (BFKL)~\cite{bfkl} 
amplitudes with energy, that would eventually lead to a violation of the 
$S$-matrix unitarity. So far, the most successful realization of the 
unitarization of BFKL amplitudes was performed within the 
Balitsky-Kovchegov (BK) framework~\cite{bk}, for the scattering of 
deeply virtual photon, $\gamma^*$, on a large nucleus. 
%in the large $N_c$ limit. 
This scheme relies on the dipole-like nature of the hard probe and 
on the large $N_c$ limit. Unfortunately, the BK formalism is not sufficient
to solve an important problem of the high energy baryon scattering.
Baryons contain at least $N_c$ constituents with a non-trivial color
connection, and consequently the large $N_c$-limit for the baryon wave 
function is much more complex than it was for the $\gamma^*$ 
(or a color dipole) projectile. This obstacle prohibited the 
direct solution of the baryon scattering problem within the 
$s$-channel approach~\cite{pr}.

Recently we attempted to analyze the baryon scattering 
amplitude within the $t$-channel framework~\cite{bm}, for $N_c = 3$. 
The formalism applied is derived in perturbative QCD and it 
resums to all orders the leading logarithmic contributions 
$(\alpha_s \ln(s))^n$ of the collision energy squared, $s$. 
%Being based on the analysis of $M \to N$ 
%scattering amplitudes in QCD, 
It generalizes the Bartels-Kwieci\'{n}ski-Prasza\l{}owicz (BKP) evolution 
scheme~\cite{bkp_bar,bkp_kp} valid for a fixed number of 
reggeized gluons in the $t$-channel, by inclusion of integral kernels 
that change the number of gluons in the $t$-channel~\cite{bkp_bar}. 
In particular, in this approach the triple-Pomeron vertex was 
obtained~\cite{jb} that was shown to match in the large $N_c$ limit the 
vertex defining the BK~equation. 
In this talk~\cite{url} results of the $t$-channel analysis of baryon
scattering amplitudes are presented: the baryon impact factors and 
their small-$x$ evolution are described.

\section{The impact factors}

The baryon impact factors for $n$~gluons are defined by multiple 
discontinuities of the elastic scattering amplitudes of baryons
mediated by $n$-gluons, see Fig.~1.
They are evaluated in the high energy limit, when only the leading
power of the large light-cone momentum component is retained.
The impact factors are given by the following formula, 
%The information about the baryon structure 
%
\beq
\label{bcal}
{B}_{n;0} (\vl_1,\vl_2,\vl_3) \; = \;
I_{qq} ^{(n)} \; \sum_{\mathrm{diagrams}}
{F}(\vl_1,\vl_2,\vl_3) \;\;
{\cal C}_n(\mathrm{diagram}),
\eeq
where the factor $I_{qq} ^{(n)} \; = \; (-ig)^n \,$ represents the 
quark scattering amplitude (without the color factor).
The overlap of the outgoing and incoming baryon wave functions
is decomposed for each diagram into the color dependent factor 
${\cal C}_n(\mathrm{diagram})$, and the momentum dependent factor, 
${F}(\vl_1,\vl_2,\vl_3)$. Note, that the impact factor depends
only on the total momentum transfers, $\vl_1$, $\vl_2$~and $\vl_3$ 
to quark lines 1, 2, and 3 respectively. 
The color part of the baryon wave function is given by the fully 
antisymmetric tensor, $\epsilon^{\alpha\beta\gamma}$. Using the
master formula (\ref{bcal}) we performed the sum over all relevant 
diagrams and evaluated the baryon impact factors up to 
four~external gluons, both in the $C$-even ($B_{n;0}$) and the $C$-odd 
($\tilde B_{n;0}$) channel.

\begin{figure}[t]
\centerline{\epsfig{file=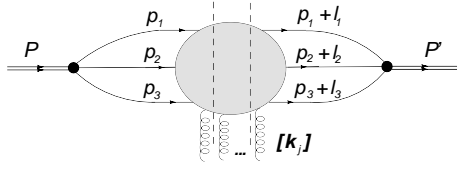,width=6cm}}
\caption{Baryon impact factor.} 
\end{figure}

For two gluons, with transverse momenta $\vk_1$ and  $\vk_2$, and 
color indices $a_1$ and $a_2$, only the Pomeron contributes. 
$B_{2;0}$ can be then represented as a sum of three pieces,
\beq
B_{2;0} (\vk_1,\vk_2)  =  {\delta\,}^{a_1 a_2}\;
\left[
\Ba_{2;0} ^{\{1,2\}} (1,2) +
\Ba_{2;0} ^{\{1,3\}} (1,2) +
\Ba_{2;0} ^{\{2,3\}} (1,2)\,
\right], 
\label{dipole}
\eeq
where we used a short-hand notation 
$\Ba_{2;0} ^{\{1,2\}} (1,2) \equiv \Ba_{2;0} ^{\{i,j\}}(\vk_1,\vk_2)$,
$\Ba_{2;0} ^{\{1,2\}}(12,0) \equiv \Ba_{2;0} ^{\{i,j\}}(\vk_1+\vk_2,0)$,
etc. In the components $\Ba_{2;0} ^{\{i,j\}}$ only quarks $i$ and $j$ 
scatter and the third quark is a spectator. All $\Ba_{2;0} ^{\{i,j\}}$ 
have the momentum structure of the color dipole impact factor, e.g.\ 
\beeq
\Ba_{2;0} ^{\{1,2\}} (1,2) & = & 
{-g^2\over 12}\; [ 
F(12,0,0) +  F(0,12,0)
- F(1,2,0) -  F(2,1,0)
],
\label{D_2,12} 
\eeeq
and similarly for $\Ba_{2;0} ^{\{1,3\}}$ and  $\Ba_{2;0} ^{\{2,3\}}$.

For three gluons in the $C$-even channel, the impact factor can be
decomposed into dipole-like components in an analogous way, 
$\,B_{3;0}\, = \, \Ba_{3;0} ^{\{1,2\}} \,+\, \Ba_{3;0} ^{\{1,3\}}\,+\, 
\Ba_{3;0} ^{\{2,3\}},$ 
and the dipole-like components have the reggeizing form known from the color 
dipole case,
\beq
\Ba_{3;0} ^{\{i,j\}} (1,2,3)  \; = \; 
{1\over 2}\,g\,f^{a_1 a_2 a_3} \; 
\left[ 
\Ba_{2;0} ^{\{i,j\}} (12,3) - 
\Ba_{2;0} ^{\{i,j\}} (13,2) + 
\Ba_{2;0} ^{\{i,j\}} (23,1) 
\right],
\label{D30}
\eeq
%indicating that a reduction 
In the odderon channel, we see a distinct color-momentum structure, 
\beq
\tilde B_{3;0}(\vk_1,\vk_2,\vk_3) \, =  \,
{d\,}^{a_1 a_2 a_3} \, E_{3;0}(1,2,3),
\eeq
where for a symmetric function $F(1,2,3)$ and for $\vk_1+\vk_2+\vk_3=0$,
$E_{3;0}$ takes the form,
\beq
E_{3;0} (1,2,3) \; =  \; 
{i g^3\over 4} \left[
2F(1,2,3) - \sum_{j=1} ^3 F(j,-j,0) + F(0,0,0) \right].
\label{3gluonodd}
\eeq

For the case of four gluons ($C=+$) we see the further emergence of the 
gluon reggeization pattern for each of the dipole-like components, 
but in addition, a new structure, $Q_{4;0}$, appears:
\beq
B_{4;0}\;=\; 
D_{4;0} ^{\{1,2\}}  \,+\,
D_{4;0} ^{\{1,3\}}  \,+\,
D_{4;0} ^{\{2,3\}}  \,+\,
Q_{4;0},
\label{B40}
\eeq
where $D_{4;0} ^{\{i,j\}}$ have the functional form known from the
color dipole case, and
\beeq
\label{eq:q40}
Q_{4;0}\, & = & 
{-ig\over 2}\; \left[\, 
{d\,}^{a_1 a_2 b}\, {d\,}^{b a_3 a_4} 
\, - \, \frac{1}{3}\, {\delta\,}^{a_1 a_2} \, {\delta\,}^{a_3 a_4} 
\right]
\left[\, E_{3;0} (12,3,4) \,+\,  E_{3;0} (34,1,2)\, \right] \;
+ \nonumber \\
& & 
{-ig\over 2}\; \left[\, 
{d\,}^{a_1 a_3 b}\, {d\,}^{b a_2 a_4} \, - \,
\frac{1}{3}\, {\delta\,}^{a_1 a_3} \, {\delta\,}^{a_2 a_4} 
\right]
\left[\, E_{3;0} (13,2,4) \,+\,  E_{3;0} (24,1,3)\, \right]\;
+ \nonumber \\
& & 
{-ig\over 2}\; \left[\, 
{d\,}^{a_1 a_4 b}\, {d\,}^{b a_2 a_3} \, - \,
\frac{1}{3}{\delta\,}^{a_1 a_4} \, {\delta\,}^{a_2 a_3} 
\right]
\left[\, E_{3;0} (14,2,3) \,+\,  E_{3;0} (23,1,4)\, \right].
\eeeq

In the case of the odderon, for four gluons, one finds that the impact factor
is fully exhausted by the reggeizing contribution, that is 
$\tilde B_{4;0}$ may be obtained from  $\tilde B_{3;0}$ by all possible 
splittings of a single gluon into two elementary gluons
with the color tensor $f^{abc}$, cf.~Eq.~(\ref{D30}).

\section{The evolution}

We demonstrated that the basic objects (modulo reggeization)
defining the baryon impact factor are the dipole-like components, 
$\Ba_{2;0} ^{\{i,j\}}$, and the functions $E_{3;0}$ and $Q_{4;0}$. 
All these functions vanish if one of the gluon transverse 
momenta vanishes. They are also fully symmetric under permutations of 
the gluon momenta (Bose invariance). Thus, they are proper initial 
conditions for the BKP evolution: $\Ba_{2;0} ^{\{i,j\}}$ for the BFKL Pomeron, 
and $E_{3;0}$ for the BKP odderon spanned by three reggeized gluons. 
$Q_{4;0}$ may be interpreted as an initial condition for a $C$-even 
three-Reggeon state, where one of the reggeized gluons has the even 
signature.

In order to analyze the structure of unitarity corrections in the 
$t$-channel approach, one has to go beyond the Reggeon number conserving 
BKP equation, and include integral kernels describing splittings of 
$2 \to n$~Reggeons. Then, the small-$x$ evolution of impact factors is
given by a set of coupled integral equations with the initial
conditions given by $B_{n;0}$ and $\tilde B_{n;0}$. 
We solved these integral equations for the baryon up to four gluons.
In the odderon channel we found only the BKP evolution preserving the 
color-momentum structure of $E_{3;0}$. In the Pomeron channel the situation
is more complex. The dipole-like pieces obey the BFKL evolution, preserving 
their color-momentum structure, but in addition a transition may occur
to a four-Reggeon state, that may be projected on two BFKL Pomerons (Fig.~2).
The amplitude of this transition is given by the $V_{2\to 4}$ vertex (related 
to the triple-Pomeron vertex), well known from the analysis of $\gamma^*$ 
scattering. In addition, we found the BKP evolution of the three-Reggeon 
state, $Q_4$. The state $Q_4$, however, may also decay into four Reggeons 
with the amplitude given by a new vertex, $W_{3\to 4}$, that may be
also be interpreted as a triple Pomeron vertex, but with the three-Reggeon
BKP Pomeron, $Q_4$, that splits into two BFKL Pomerons, see Fig.~2. 
We point out that the possible direct two-Pomeron coupling to the baryon 
was not found. The lack of the direct two-Pomeron coupling, however, 
essentially relies on taking in account only the lowest Fock component 
of the baryon, and it holds only in the leading logarithmic $\ln(s)$ 
approximation.

\begin{figure}[t]
\leavevmode
\begin{center}
\epsfig{file=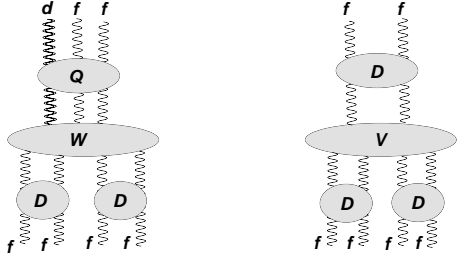,width=6.0cm}\\
\end{center}
\caption{Transition vertices $W_{3\to 4}$ (left) and $V_{2\to 4}$ (right); 
reggeized gluons with the odd ($f$) and the even ($d$) signature are 
indicated.}
\end{figure}

\section{Conclusions}

We have analyzed the high energy scattering of a baryon projectile.
The baryon, represented by three constituent quarks,
was found to couple to the BFKL Pomeron, the BKP (three-Reggeon) 
odderon and a new state, a BKP Pomeron spanned on three Reggeons,
out of which one has an even signature. The BFKL Pomeron may couple to
one of dipole-like pieces of the baryon. Each dipole-like component of 
the baryon has the color-momentum structure of the genuine color dipole.
The evolution of those states was analyzed up to four reggeized gluons 
in the $t$-channel. The dipole-like components were found to evolve in the
same way as the color dipoles. Specifically, their evolution is driven
by the BFKL equation, followed by a possible splitting of the BFKL Pomeron
into four reggeized gluons (two Pomerons). The three-Reggeon Pomeron obeys
the BKP equation and it may split into four reggeized gluons. This 
transition is driven by a new $3\to 4$ reggeized gluon vertex.

% ****************************************************************************
% BIBLIOGRAPHY AREA
% ****************************************************************************

\begin{footnotesize}
% IF YOU DO NOT USE BIBTEX, USE THE FOLLOWING SAMPLE SCHEME FOR THE REFERENCES
% ----------------------------------------------------------------------------

\end{footnotesize}

% ****************************************************************************
% END OF BIBLIOGRAPHY AREA
% ****************************************************************************

\end{document}